# MIMO Cluster Cooperative Assignment Cross Layer Scheme for Hybrid Wireless Sensor Networks

Yingji Zhong
UWB Wireless Communication Research Center, Inha University, Korea
School of Information Science and Engineering, Shandong University, China
zhongyingji32@sdu.edu.cn

Kyung Sup Kwak
UWB Wireless Communication Research Center
Inha University
Incheon, 402-751, Korea
kskwak@inha.ac.kr

Sana Ullah

*Abstract*—The dual-cross scenario of the hybrid wireless sensor networks (WSNs) is studied and a novel MIMO Cluster Cooperative Assignment Cross Layer Scheduling Scheme (MCCA-CLSS) is proposed in this paper. The comparison and the predominance of the proposed scheme are demonstrated, the clusters are optimized. With the help of the simulations, the relative energy consumption and the end-to-end blocking probability are all improved. The addressing ratio of success in the condition of the unchanged parameters and external information can be increased and the network can tolerate more hops to support reliable transportation by the proposed scheme.

## I. INTRODUCTION

Due to the limited energy and difficulty to recharge a large number of sensors, energy efficiency and maximizing network lifetime have been the most important design goals for wireless sensor networks (WSNs). However, channel fading, interference, and radio irregularity pose big challenges on the design of energy efficient communication and routing protocols in the WSNs. As the MIMO technology has the potential to dramatically increase the channel capacity and reduce transmission energy consumption in fading channels [1], cooperative MIMO schemes have been proposed for WSNs to improve communication performance [2]. In those schemes, multiple individual single-antenna nodes cooperate on information transmission and/or reception for energy-efficient communications.

In the cooperative process, the secondary cooperative terminals transmit the signals on the frequency band assigned to the primary system by sensing the radio frequency band in order to avoid the interference toward the primary systems. However, it is difficult to recognize the status of the frequency band when the primary terminals only receive the signals. Therefore, the proposal of the MIMO cooperative scheduling in order to realize a wide area secondary communication system by using multi-hop networks is necessary. Although the large transmission power on single-hop networks can support the large communication area, the interference toward the primary system also becomes large if the primary system exists between the primary transmitter and the receiver. In the node cooperation, the power of each node is suppressed to minimize the interference toward the primary system and the area of communication can expand by using the multi-hop networks[3].

The cooperative assignment problem[4] seeks an assignment of the fewest channels to a given set of the nodes with specified transmission ranges without any primary collision or secondary collision. It is a classic and fundamental problem in wireless hybrid WSNs. It is NP-hard even when all nodes are located in a plane and have the same transmission range. Therefore, only polynomial-time approximation schemes can be expected. The performance of a polynomial-time approximation algorithm is measured by its approximation ratio, which is the supreme, over all instances, of the ratio of the number of channels output by the schemes to the minimum number of the channels. However, the schemes are complicated because the location and the active time of the primary system are not fixed. To our assumption, the hybrid WSN networks has the potential to enable a large class of applications ranging from assisting elderly in public spaces to border protection that benefit from the use of numerous nodes that deliver packets. In multi-hop wireless networks, there is a strict interdependence cross layer coupling of functionalities among functions handled at all layers of the communication stack. Multiple paths may exist between a given source-sink pair, and the order of packet delivery is strongly influenced by the characteristics of the route chosen.

In order to improve the robustness of the node cooperation without complicated scheduling framework, we propose a novel cross layer MIMO cooperative scheduling scheme for the hybrid WSN networks in this paper. Addressing the existing questions and designing a viable end-to-end solution may be the first attempt. We also identify key design parameters and present a methodology to optimize cross layer efficiency, data quality and coverage area. To the best of our knowledge, such a study has not been thoroughly conducted in [5] and [6]. To this end, the scheme we proposed can support packet-based delay guarantees that must be delivered with a given probability.

The rest of the paper is organized as follows. In Section II, we give the evaluation architecture. In Section III, we propose the cross layer MIMO cluster cooperative assignment scheme. In Section IV, we evaluate the performance of the proposed scheme and analyze the improvement of the cooperative scheduling guarantee via simulation. Finally we give the conclusion in Section V.

## II. EVALUATION ARCHITECTURE

In randomized cooperation, each node projects the rows of the state matrix can generate a randomized state $\tilde{x}_r = Xr_r = G_{M \times N}(s)r_r$. As the assumption of squared power path loss, it can be modelled by

$$E_t(1) = \frac{P_{ct} + JP_{cr}}{B} - 2(1+\alpha)N_f \sigma^2 \ln(P_b) G_1 e_{max}^2 M_i \quad (1)$$

where $e_{max}$ is the maximum distance from the cooperative nodes to the cluster head, $\alpha$ is the efficiency of the RF power amplifier, $\sigma$ is the gain factor at $e_{max} = 2m$, $M_i$ is the link margin, $N_f$ is the receiver noise figure, and $P_{ct}$ and $P_{cr}$ are the circuit power consumption of the transmitter and receiver, respectively. The diversity that can be obtained through this scheme depends on the statistics of the resulting equivalent states and on the particular selection of the state $G_{M \times N}(s)$ just as it does for the deterministic assignment. For simplicity and to gain intuition, we consider the transmission model where the channels between the source node and the destination node are orthogonal and $i=1,2,3,4$. A message that contains a request for cooperation is stored in the relay buffer, whose transmission is synchronized by the preamble sequence received in the message containing the request. The state parameters in network layer need to be informed about the state of the relay buffer. In general, the half-duplex constraint of the transmission model mandates that the destination node be inactive when the source node is busy, but the upper layer can also prevent cooperative transmission for it. The average energy consumption per-bit transmission by BPSK in such a scheme can be approximated by

$$E_t(2) = \frac{JP_{ct} + P_{cr}}{B} + (1+\alpha)\frac{N_0}{P_b^{1/J}} \sum_{j=1}^{J} \frac{16\pi^2 e_{jt}^{k_{jt}}}{\lambda^2 H_t H_r} \sigma^2 M_i N_f \quad (2)$$

where $N_0$ is the single-sided noise power spectral density, $P_b$ is the desired BER performance, $\lambda$ is the carrier wavelength, $H_t$ and $H_r$ are the transmitter and receiver antenna gains, respectively, The sum capacity must be characterized in terms of maximization as

$$R^{coop} = \max_{T_r,(Q_1+Q_2) \leq P} \log \left| I + \tilde{M}_1^T N_s \tilde{M}_1 + \tilde{M}_2^T N_r \tilde{M}_2 \right| \quad (3)$$

where the maximization is over covariance matrices $Q_1$ and $Q_2$, with $\tilde{M}_1 \triangleq \begin{bmatrix} M_1 \\ \beta M_2 \end{bmatrix}$, and $\tilde{M}_2 \triangleq \begin{bmatrix} \beta M_1 \\ M_2 \end{bmatrix}$. For fixed $P_t$ and $P_r$, the achievable minimum rate is $\min(2R_t, R_{coop})$.

The evaluation model is the multiple-cell environments with seven cells as shown in Fig.1, in which the nodes are in point wise uniformity. Analysis is based on two-dimension scenario, that is to say, the nodes and the base stations are on a cross. The nodes are placed at every $D$ distance unit from base stations and define intervals $[nD, (n+1)D]$ of length $D$ on the Cross $C_s(N)$. The base station of Cell 1 is on the middle of $C_s(N)$. Assume $C_s(N)$ is locally compact space. Let $\cup A=N$, for each $n \in N$, define $F_n=\{1,2,\ldots,n\}$ and $U_n = \{\{A\} \cup (A/F_n) : A \in \{x\} : x \in F_n\}$ then $U_n$ is the open covering of $C_s(N)$. For each $x \in C_s(N)$, when $x \in F_n$, $st(x,U_n) = \{x\}$, and $x \in A$, $st(x,U_n) = \{x\} \cup (x, \tilde{y}/F_n)$. So $\{U_n\}$ is the development of $C_s(N)$. Thus, $C_s(N)$ is the developable space of locally compact. Let $\mathcal{K}$ is the compact subspace of $C_s(N)$, since $A$ is the closed discrete subspace of $C_s(N)$, so $\mathcal{K} \cap A$ is the finite set, $\mathcal{K}$ is countable set of $C_s(N)$ and $\mathcal{K}$ can be metric, it is obvious that $\mathcal{K}$ has countable neighborhood basis in $C_s(N)$. $C_s(N)$ is a dividable space, if it has point-countable basis, then $A$, the subspace of $C_s(N)$ has countable basis, this is contrary to reality that $A$ is the uncountable closed discrete subspace of $C_s(N)$. So $C_s(N)$ has no point-countable basis. As for the multi-scenario shown in Fig.2, the relays are used for exchanging control messages and assigning the dedicated channel. In high bandwidth applications, the use of a separate channel for channel arbitration alone does not allow best utilization of the network resources. It is necessary to directly maximize the achievable rates over all choices of $P_r$, where the same channel is used for both data and channel arbitration. The scaling term $\beta$ can be made close to one. Thus the composite channels capacity is equal to the point-to-point MIMO capacity of the original channel [7,8]. Such model undoubtedly improve bandwidth efficiency and introduce the problem of distinct channel assignment and need to account for the delay to switch to a different channel as its cumulative nature at each hop affects flows. Only when a node detects the overloaded channel condition that it concerns is the adjustment candidate selection process triggered. It checks the feasibility of possible candidates on the most overloaded channel under the premise that it can directly benefit from the adjustment. So it actually selects from only a few possible choices and does not cause the cardinality explosion of the candidate sets. In the multiple cell environment, the evaluation model with multi-scenario as shown in Fig.2 on which the nodes are in point wise uniformity. It can be seen that the multi-scenario of the cross topology belongs to the combination of two Gillman-Jerison spaces [5,9]. However, a distinct difference in this case is that the nodes themselves can also function as active mobile nodes. These nodes under group-oriented operation are capable of initiating communications not only with their mobile nodes, but also with others. The attributes of Gillman-Jerison Space are potentially worthy for the novel scheme proposal.

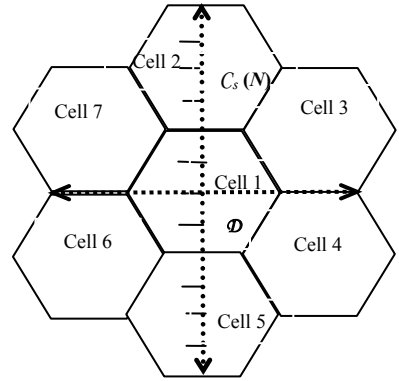

Fig. 1. The scheme of cross topology scenario

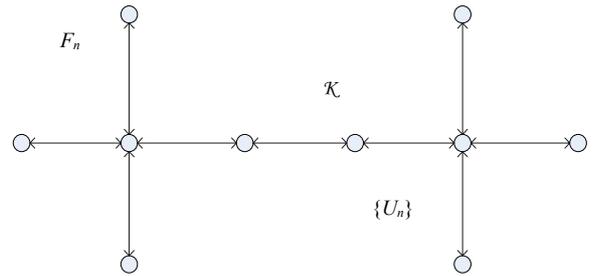

Fig.2 The multi-scenario of the cross topology

The candidate receivers evaluate the channel condition based on the physical layer analysis of the received RTS message. If the channel quality is better than certain level, the given receiver is allowed to transmit CTS. To avoid collision when two or more intended receivers are qualified to receive data, a service rule is applied. The listing order of intended receivers in the RTS announces the priority of the MAC among the candidate receivers. The closer the receiver address to the top of the receiver list, the higher the priority to access media[10]. To prioritize the receivers, different $N_s$ are employed. The receiver with highest priority among those who have capability to receive data packet would reply CTS first. Since all candidate receivers are within one-hop transmission range of the sender and the carrier sensing range are normally larger than two hops of transmission range, the CTS should be powerful enough for all other qualified candidate receivers to sense. These receivers would yield the opportunity to the one transmitting CTS in the first place and the $j_{prio}$ with the good channel condition remains highest priority. The maximum candidate receiver list is scheduled if there are enough data packets targeting it. Longer receiver list means more diversity can be exploited, but also means the waiting time would increase before the transmitter can make sure there is no qualified receiver. Assume that the probabilities for candidate receivers to successfully receive the intended data packet are identical and independent. During the simulation, we find it already yields significant throughput gains under the 8 flows condition. Multicast RTS and prioritized CTS with channel awareness parallelize the multiple serial RTS/CTS messages, so the overhead of channel contention and channel probing can be reduced. More importantly, the cooperative scheduling can be alleviated and accounted for a dynamically changing topology due to nodes dying off or new ones being added. For simplicity, we assume that the interference range of each node is almost equal to its transmission range in this specific scenario. All the links are free of transmission error and the raw capacity of each link is 256Kbps. The node chooses the multi-hop transmission through neighbors with the help of the fixed selection criterion rather than access the base stations directly. Since scheduling is going to be used at the base stations, it is possible to assign different carrier frequencies to the different multi-hop routes. With the help of $G_1$, the multi-scenario can be induced. Consider the optimal cell tessellation, the cooperative diversity based on multi-hop throughout the network using the centers of the cells and compare the total power spent. Each node makes decision independently in the hybrid networks and causes inconsistence when two nearby nodes adjust their topology simultaneously[11,12]. Thus it is required that both the channel adjustment and power adjustment should guarantee the exclusiveness of the cross layer adjustment in the interference area of the node. Under this premise, each node may locally make adjustment decisions without considering the disturbance of the neighboring nodes. The WSNs we considered is denoted by $G = (V, E)$ where $V$ is the set of nodes and $E$ is the set of links. Let $G_1 = (V, E')$ represent the graph induced by the scheme and $G_2 = (V, E'')$ represent the graph induced by the channel assignment scheme where $G_1 \subseteq G_2 \subseteq G$. If the network is set to meets the same SNR threshold constraints, each node has to transmit $P_r \geq \tau R^2$.

### III. CLUSTER COOPERATIVE SCHEDULING SCHEME

If the nodes result in an action profile where each user's action is a best response to the others in the cooperative scheduler module, the Nash equilibrium is reached[13]. In other words, the Nash equilibrium is the action profile where no user has an incentive to deviate by choosing another action given that the other user's action is fixed. Formally, the Nash equilibrium can be acquired by the following action profiles for each node. The primary concern in our scheme is accomplished at the cost of latency and by allowing throughput degradation. A sophisticated duty cycle calculation based on permissible end-to-end delay needs to be implemented and coordinating overlapping listen period with neighbors based on this calculation is a difficult research challenge. When the end-to-end traffic can be split in the dual-cross scenario, the number of the routes between source and destination should be two or more. That is to say, the flow going through the route is no longer an integer and the traffic demands can split.

The novel MIMO Cluster Cooperative Assignment Cross Layer Scheduling Scheme (MCCA-CLSS) is shown below.

INPUT: the set of the nodes and $\beta, P_t, P_r$ in the Euclidean plane and positive real constants representing the communication ranges of the nodes, respectively.

OUTPUT: the set $U_n = \{u_1, u_2 \ldots u_n\}$ of the open covering in the multi-scenario.

```
Init( )
{
  let G = (V,E) be undirected edge-weighted complete graph on the vertices
    in U_n , where for 1 ≤ i < j ≤ n , the weight of the state set (x_ri, x_rj) is
    V(x_ri, x_rj) = ||x_ri x_rj||.
  maximize U_s(p_s, n) without constraints;
  maximize U_r(p_r, m) ;
  compute a minimum spanning tree F_n of G
  analyze the contention of links on channel c in two hop range;
    if i is bound to nodes of neighboring cluster then
      assign i, the channel assignment from its neighbor assignment;
    else
      set  y_ij^td = y_ji^td  and  i ≠ 0  iteratively  update  q^(τ_i)
      as q_l^(τ_i+1) = μ_l/t_l^(s) - 1/M_ll (M_lj q_j^(τ_i) + σ_l^2) ;
  project q_l^(τ_i+1) into power constraint interval [0, q_{l,max}] ;
  calculate the resource assignment for channel c;
    end if
  calculate R^coop on channel c and corresponding priority for each group;
  repeat until q^(τ_i) converges;
  set q^(s+1) = q^(τ_i) and refresh G_{M×N}(s) , E_t(1) and E_t(2) ;
    if no channel overloaded
       return;
    end if
    if feasible
       select adjustment candidate with R^coop and begin negotiation;
    end if
```

```
    if (p_s^*, n^*, p_r^*, m^*) does not change
        Rate s_n E_t(1) and E_t(2)
    end if
for state set (x_ri, x_rj) ∈ F_n
    if 0 < ||x_ri x_rj|| < r then
        place one cognitive node g_r at the midpoint of the line segment
[x_ri, x_rj]
    else if r ≤ ||x_ri x_rj|| then
        place two cognitive nodes, g_ri, g_rj on the line segment [x_ri, x_rj]
end if
for j ∈ N_s
    if Prand(N_s) < W(i) and priority of t is not Φ
        Recover β, P_t, P_r and N_s;
    end if
    if Load(i) < N_r
        then select adjustment candidate with minimal load and begin
        distribution;
    end if

    analyze the contention of links on channel c in two hop range;
    if i is bound to nodes of neighboring cluster
        assign i, the channel assignment for the next assignment;
    end if
end
}
```

In our scheme, the non-forwarding action is always in Nash equilibrium. The topology construction is performed during the network initialization phase when no user traffic is present in the network. To fully reduce the co-channel interference and achieve higher gains of network performance, the topology attributes and power constraints should be jointly considered to exploit not only channel diversity but also spatial reusability. Firstly, we sort all the node pairs in ascending order according to their minimum distance. Secondly, $R^{coop}$ evaluation runs on every node in the network to check whether the flow can be all routed or not. Coordinating the sleep-awake cycles between neighbors is generally accomplished though MIMO scheduling exchanges. In case of dynamic duty cycles based on perceived values of instantaneous or time averaged end-to-end latency, the overhead of passing frequent schedules also needs investigation in light of the ongoing high data rate message. The negotiation mechanism is indispensable to make sure all the affected nodes are consented with the adjustment. The initiator sends an adjustment preparation broadcast packet to notify the participator and onlooker nodes, which consequently reschedule their channel check timers to ensure no new negotiation happens. The node sends adjustment request that includes the adjustment information of all affected nodes to the participator nodes, and then starts the reply waiting timer and waits for their reply. After receiving the adjustment request, participators check the feasibility for the required adjustment and then send an adjustment reply packet to the initiator about their decision. Besides, the participators broadcast an adjustment preparation broadcast packet to notify its onlooker with the same purpose. According to our scheme, the transmit power of the node to communicate with the cluster head can be described by $P_{jt} = G(d_{jt}, k_{jt}) \frac{N_0 B}{P_b^{1/j}}$ When the reply waiting timer expires, the initiator checks the adjustment replies to see if all the participators agree with the adjustment, If yes, the initiator sends an adjustment notification to all the participator and starts the notification timer. If not, the negotiation fails. When the notification timers expire, both the weight and the state set make corresponding adjustment. Until now, the adjustment really takes effect. Finally, the initiator and participators broadcast their adjustment information through which onlookers could update their two-hop neighbor information timely. In order to avoid packet loss during negotiation, adjustment request, adjustment reply, adjustment notification packet are sent three times sequentially to ensure reliability. If the adjustment of a node is related with route change, there is an additional process of route negotiation. When it receives a packet with the obsolete path information, the node tunnels the packet to the up-to-date route. Meanwhile it sends a route update packet to the traffic source. The traffic source node updates its cache and then sends a route accept packet back to the node. Only when all routes accept packets are received does the node remove the route change information. The operation should be terminated when the transmission power reaches to maximum. In this scheme, the topology and power consumption of each node can be optimized due to the minimum link occupation. The power update is the best response of link player given the tax rate and assessment of others' action. As for the tax rates converge, it can be induced to a stable Nash equilibrium. Such equilibrium strikes a balance between minimizing interference and maximizing rate.

IV. SIMULATIONS AND DISCUSSION

The terrain model we used is a 12*km*×6*km* rectangular area with dual-cross in the multi-scenario. In each cross, the nodes are pseudo-randomly moving along the cluster cells under *NS-2*. All the links between nodes are bi-directional. Each cell has a base station with omni-directional antenna at the center point and its radius is 3*km*. Each node can support 128 available data channels. As for handoff mechanism, hard handoff was used in the evaluation model and connectivity is considered under Poisson Boolean Model in this kind of sparse network. We use 64 TCP flows in the multi-scenario and the simulation time for each point is 4096s. We assume that the power consumption is based on the distance from the transmitting nodes to the destinations. Employing the proposed scheme, the relative energy consumption and the end-to-end blocking probability are examined in different number of the nodes. Shown in Fig.3 and Fig.4 respectively. As expected, the use of the proposed scheme can optimize the available channel capacity and the relative performance.

Fig.3 gives the relative energy consumption with varying number of the nodes. As the relative gain increases, the achievable rates increase accordingly. The system with cooperative scheduling scheme performs better than the one without MCCA-CLSS, which only outperforms strategy game. For large number, the ratio approximates 1.6, hence the gain in total energy consumption for the reliability balancing strategy is 49.15%. More important and more significant is the gain in network lifetime, which is determined by the lifetime of the current node. Notice that the two curves are independent of the channel states because they assume perfect condition. The proposed scheme is virtually identical to the MIMO cooperative scheduling until the point where the power gain comes very close to the real utility. Thus, it seems that is not necessary to do cooperation in the proposed scheme especially when the gain is interrupted by the addressing ratio and the permitted hops.

Fig.4 gives the end-to-end blocking probability with varying number of the nodes. Observe that when the node number is less than 7000, the probability of the end-to-end blocking is quite large because the throughput and the congestion is actually in idle state. This is reasonable because no multi-user diversity gain can be achieved in case there is only one user has longer scheduling time than that of MAC. When the number of the nodes increases, the throughput gain benefited from opportunistic scheduling starts to show. When the number of flows increases to 8000 or above, the probability of the end-to-end blocking exceeds 0.07% and the gain maintains relatively stable. The minimal optimization is 29.55% when the number of flows reaches to 12000. What is more, the addressing ratio of success in the condition of the unchanged parameters and external information can also be increased. The reason is that the probability of all candidate receivers is not satisfied to receive a packet at any given time is very low. When the number of flows goes up, almost each time access point sends an RTS and receives CTS to continue data delivery.

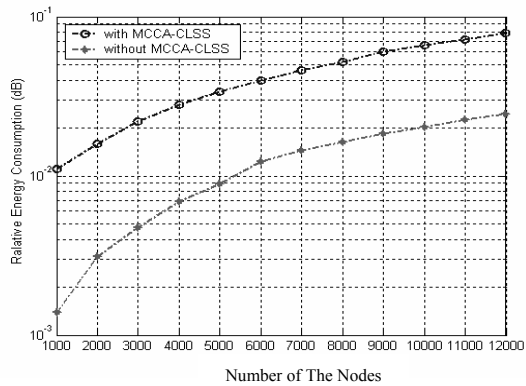

Fig.3. Relative energy consumption with varying number of the nodes

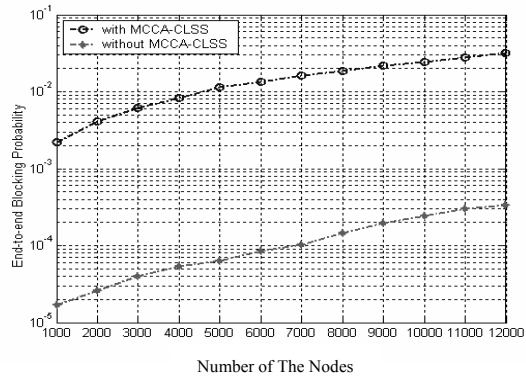

Fig.4. End-to-end blocking probability with varying number of the nodes

## V. CONCLUSION

In this paper, a novel MIMO cluster cooperative scheduling scheme with the dual-cross scenario in cross layer aspects was proposed. The comparison and the predominance of the proposed scheme are demonstrated by the simulation results analysis. The relative energy consumption and the end-to-end blocking probability are improved with the help of the simulations. The addressing ratio of success in the condition of the unchanged parameters and external information can be increased and the network can tolerate more hops to support reliable transportation by the proposed scheme.

ACKNOWLEDGMENT

This research was supported by the MKE(The Ministry of Knowledge Economy), Korea,under the ITRC(Information Technology Research Center) support program supervised by theIITA(Institute for Information Technology Advancement)(IITA-2009-C1090-0902-0019) and the Natural Science Foundation in Shandong Province(Q2007G01).